\documentclass[pdflatex,sn-mathphys-num]{sn-jnl}

\usepackage{graphicx}%
\usepackage{multirow}%
\usepackage{longtable}
\usepackage{amsmath,amssymb,amsfonts}%
\usepackage{amsthm}%
\usepackage{mathrsfs}%
\usepackage[title]{appendix}%
\usepackage{xcolor}%
\usepackage{textcomp}%
\usepackage{manyfoot}%
\usepackage{booktabs}%
\usepackage{algorithm}%
\usepackage{algorithmicx}%
\usepackage{algpseudocode}%
\usepackage{listings}%
\usepackage{comment}
\usepackage[colorinlistoftodos]{todonotes}

\usepackage{hyperref}

\usepackage{soul} 

\theoremstyle{thmstyleone}%
\theoremstyle{thmstyletwo}%

\theoremstyle{thmstylethree}%

\begin{document}

\title[Article Title]{Automatic techniques for security bug report identification: A systematic mapping study}


\author{\fnm{Muhammad} \sur{Laiq} and \fnm{Usman} \sur{Nasir}}\email{muhammad.laiq@bth.se, usman.nasir@bth.se}

\affil{Blekinge Institute of Technology, Karlskrona, Sweden}

\abstract{
Several studies have investigated the use of automated techniques to help practitioners identify security bug reports early in the defect management process. However, currently there is no comprehensive overview of this research area. Such an overview will help identify future research directions and consolidate the potentially relevant existing solutions. This study aims to provide a comprehensive overview of automated techniques for identifying security bug reports. To achieve this, we conducted a systematic mapping study and identified 28 primary studies.

The study's findings show that the literature has evaluated various automatic techniques for identifying security bug reports, including traditional machine learning, deep learning, and more advanced large language models. The two main challenges identified are the lack of labeled data and data imbalance. Furthermore, we observe that existing studies lack evaluations in real industrial settings and rely mostly on archival data from open-source repositories. The most commonly used data sources in these studies include Wicket, Ambari, Camel, and Derby. Overall, the literature appears fragmented, highlighting the need for future work to focus on comprehensive evaluations of the most promising techniques.
}

\maketitle

\section{Introduction}
Bug tracking systems, such as Bugzilla\footnote{https://www.bugzilla.org/} and Jira\footnote{https://www.atlassian.com/software/jira}, receive a large number of bug reports from various stakeholders, including end users, developers, and testers. These reports are heterogeneous in nature and include a range of problems, such as functional defects, performance issues, usability concerns, and security vulnerabilities. Among these categories, security-related bug reports are important; failure to identify them in a timely manner can expose software systems to exploitation.

Manually reviewing incoming bug reports to identify security-related issues is laborious and error-prone \cite{wu2021data}. Moreover, as the volume of incoming reports continues to grow in large-scale software systems, the manual approach becomes increasingly infeasible and does not scale effectively with operational demands. Thus, to assist practitioners in detecting security bug reports early in the defect management process, researchers have proposed a variety of automated techniques (for example, \cite{PS16,PS18,PS25,PS6,PS3}). These techniques range from traditional text mining approaches to more advanced machine learning (ML) and deep learning methods. 

Despite the growing number of studies on automated identification of security bug reports, there is currently no comprehensive overview that organizes and structures this body of work. A comprehensive overview of automated techniques for identifying security bug reports is essential as it will guide future research and support informed decision-making in practice. For example, it would help practitioners identify suitable techniques for adoption in practice. Similarly, researchers can be assisted in identifying gaps and underexplored areas, as well as in observing emerging trends.

In response, we conducted a systematic mapping study \cite{petersen2008systematic} to identify primary studies focused on identifying security bug reports, answering the following overarching research question: \textbf{\textit{RQ: Which automatic techniques have been proposed in the literature for identifying security bug reports?}} 

Following a systematic approach, we developed a comprehensive map that covers various aspects, including the used techniques, datasets, features, and evaluation metrics. The main contributions of this mapping study are as follows.
\begin{itemize}

    \item A comprehensive overview/map of existing work on automated techniques for identifying security bug reports.
    
    \item Several insights from the existing literature on identifying security bug reports. For example, challenges such as class imbalance and the limited effectiveness of GPT-based models in this task, the continued strength of traditional ML techniques, such as Logistic Regression, as competitive baselines, and the promise of emerging approaches, such as SetFit, for achieving high accuracy.
    
    \item Several potential future research directions for identifying security bug reports. For example, the need to evaluate the proposed techniques in real industrial settings and to investigate other potentially relevant aspects beyond prediction accuracy, such as their explainability.
\end{itemize}

This paper is structured as follows. Section \ref{sec:research-method} describes the study research method, i.e., search strategy, selection criteria, data extraction, data analysis and classification, and threats to validity. The results are presented in Section~\ref{sec:results}. Section~\ref{sec:discussion} discusses the overall findings of the study, while Section~\ref{sec:conclusion} concludes the paper.

\section{Research method} \label{sec:research-method}
\begin{figure}[!ht]
    \centering
    \includegraphics[width=\columnwidth]{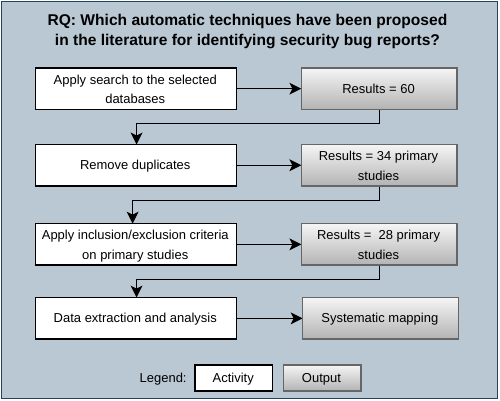}
    \vspace{-4mm}
    \caption{Overview of the systematic mapping process}
    \label{fig:sms-process}
\end{figure}

We conducted a systematic mapping study \cite{petersen2008systematic} to answer the research questions listed in Table \ref{tab:research-questions}. Figure~\ref{fig:sms-process} provides an overview of the systematic mapping process.

\begin{table*}[ht]
    \centering
    \small
    \caption{Research questions}
    \label{tab:research-questions}
    \begin{tabular}{p{12.5cm}}
    \toprule
     \textbf{Research question and its sub-research questions} 
    \\ \midrule 
   \textbf{RQ: Which automatic techniques have been proposed in the literature for identifying security bug reports?} With this research question, we aim to identify evaluated automatic techniques. In the following research questions, we further collect information such as the evaluation metrics, datasets, and evaluation approaches used. \\
    ---\textbf{RQ1: Which automatic techniques have been proposed?}\\ 
    ---\textbf{RQ2: Which datasets have been used?} \\
    ---\textbf{RQ3: Which features have been used?} \\
    ---\textbf{RQ4: Which pre-processing techniques have been used?} \\
    ---\textbf{RQ5: Which evaluation metrics have been used?} \\ 
    ---\textbf{RQ6: Which evaluation approaches have been used?}
     \\ \bottomrule
    
\end{tabular}
 \end{table*}  

\subsection{Search strategy}\label{sec:search-strategy}

Table \ref{tab:db-hits} presents the search string and the number of papers found in the selected databases on the topic. Our search focuses on identifying papers that classify bug reports into security and non-security categories. 
In addition, we also aim to identify papers that may not explicitly discuss the classification of bug reports but use terms such as `predict', `determine', or `identify' in the context of security bug reports. It should be noted that we do not focus on covering the literature on general bug report classification \cite{laiq2026automatic}, i.e., classifying bug reports as valid/invalid \cite{laiq2024industrial} or other types such as bug, question, and feature \cite{laiq2023intelligent}.

As shown in Table \ref{tab:db-hits}, the search was performed in ACM, Scopus, Web of Science, and IEEE Xplore (on March 23, 2026). The search was conducted over titles, abstracts, and keywords in ACM, Scopus, and Web of Science. In IEEE Xplore, a full-text search was performed using the command search feature.

\begin{table}[ht]
    \centering
    \caption{Search string and its results in the selected databases}
    \label{tab:db-hits}
    \begin{tabular}{p{4cm}p{3cm}}
    \toprule
        \multicolumn{2}{p{12cm}}{\textbf{Search string:} ((``bug report classification" OR ``defect report classification" OR ``issue report classification") OR (``classifying bug report" OR ``classifying defect report" OR ``classifying issue report") OR (predict OR determine OR identify)) AND (``security bug report"))}
    \\  \midrule 
     \textbf{Database} & \textbf{Search results} 
    \\  \midrule 
    Scopus & 33
    \\  \midrule
    
    Web of Science & 14
    \\ \midrule
    
    IEEE Xplore & 12
    \\  \midrule
    
    ACM & 1
    \\ \midrule    
     & \textbf{Total = 60}
    \\ \bottomrule

    \end{tabular}
\end{table}

\subsection{Study selection and outcome}\label{sec:selection-criteria}

We selected papers based on their titles and abstracts, reading the full text only when necessary to make a decision. The study selection criteria were as follows:

\vspace{2mm}
\noindent\textbf{Inclusion criteria}
\begin{itemize}
    \item [] \textbf{I1:} The paper is peer-reviewed.
    \item [] \textbf{I2:} The paper is written in English. 
    \item [] \textbf{I3:} The paper focuses on identifying security bug reports using an automated technique.
\end{itemize}

\noindent\textbf{Exclusion criteria} 
\begin{itemize}
    \item [] \textbf{E1:} The paper is not peer-reviewed
    \item [] \textbf{E2:} The paper is not written in English.
    \item [] \textbf{E3:} The paper does not focus on the identification of security bug reports using an automated technique.
\end{itemize}

Figure \ref{fig:sms-process} shows the number of included and excluded papers at each stage. Of the initial 60 papers, 26 duplicates were identified, leaving 34 unique papers. The selection criteria led to 28 papers being chosen for data extraction.

In total, 6 papers were excluded according to the exclusion criteria mentioned above. Three papers were excluded because they did not focus on identifying security bug reports. The other three papers were not written in English.

\subsection{Data extraction and analysis}\label{sec:data-extraction}

We extracted the following data from the selected primary studies:
\begin{enumerate}
    \item Meta information, e.g., title, year, and publication type.
    \item Proposed techniques \textbf{(RQ1)}, e.g., Logistic Regression and Random Forest.
    \item Datasets used \textbf{(RQ2)}.
    \item Features used \textbf{(RQ3)}, e.g., description and title of bug reports.
    \item Pre-processing techniques used for feature extraction \textbf{(RQ4)}, e.g., TF-IDF (Term Frequency-Inverse Document Frequency) and BERT-based tokenizer.
    \item Evaluation approaches \textbf{(RQ5)}, e.g., 10-fold cross-validation and test on a fixed test set.
    \item Evaluation metrics \textbf{(RQ6)}, e.g., Accuracy, F-score, Precision, and Recall.
\end{enumerate}

The extracted data were presented in tables following their classification and analysis. We categorized the proposed techniques into high-level groups, including traditional ML, deep learning, and large language models (LLMs). In addition, the studies were organized by evaluation approach, evaluation metrics, and pre-processing techniques.

\subsection{Validity threats}\label{sec:validity-threats}

\textbf{Missing relevant studies:} A potential threat to this mapping study is the omission of relevant studies on the topic. To mitigate this, we carefully designed our search string to identify relevant literature on the automatic identification of security bug reports. Additionally, we have used two major indexing (i.e., Scopus and Web of Science) and two publisher-specific databases (IEEE Xplore and ACM), without restricting publication year, to ensure broad coverage. These databases adequately cover software engineering literature \cite{kitchenham2016evidence,usman2023quality}.

\noindent\textbf{Bias in paper selection:} Another potential threat to this mapping study is bias during the paper selection process. To mitigate this, the second author reviewed all the excluded papers; we found no disagreements.

\noindent\textbf{Data extraction validity:} Another potential threat is bias during data extraction. To mitigate this threat, two reviewers participated in the data extraction process. The first author extracted data from all 28 studies, while the second author reviewed five randomly selected primary studies to verify the correctness of the extracted data. No discrepancies were found between the two authors.

One limitation is that we did not perform quality assessments of the primary studies included in this study. However, this is not a major threat to the results, as mapping studies are exploratory in nature, aiming to provide a structured overview of a research area, identify trends, and highlight gaps, rather than assess the strength or rigor of the evidence \cite{petersen2015guidelines}.

\begin{table}[!ht]
    \centering
    \scriptsize
    \caption{List of included primary studies}
    \label{tab:included-PSs}
    \begin{tabular}{p{0.8cm}p{11.5cm}}
    \toprule
    \textbf{Id} & \textbf{Title}
    \\ \midrule
PS1 & An Empirical Study to Investigate Class Imbalance Issue for Improving Security Bug Report Classification Prediction \cite{PS1} \\

PS2 & A bug Mining tool to identify and analyze security bugs using Naive Bayes and TF-IDF: A Comparative Analysis \cite{PS2} \\

PS3 & GPTs are not the silver bullet: Performance and challenges of using GPTs for security bug report identification \cite{PS3} \\ 

PS4 & Security and performance bug reports identification with class-imbalance sampling and feature selection \cite{PS4} \\

PS5 & Data Detection in Wireless Sensor Network Based on Convex Hull and Naïve Bayes Algorithm \cite{PS5} \\

PS6 & Text Filtering and Ranking for Security Bug Report Prediction \cite{PS6} \\

PS7 & A Comparative Study of Class Rebalancing Methods for Security Bug Report Classification \cite{PS7} \\

PS8 & Sais: Self-Adaptive Identification of Security Bug Reports \cite{PS8} \\

PS9 & LTRWES: A new framework for security bug report detection \cite{PS9} \\

PS10 & Improving the Security of Wireless Network through Cross-project Security Issue Prediction \cite{PS10} \\

PS11 & CASMS: Combining clustering with attention semantic model for identifying security bug reports \cite{PS11} \\

PS12 & Locality-based security bug report identification via active learning \cite{PS12} \\

PS13 & A New Method of Security Bug Reports Analysis \cite{PS13} \\

PS14 & Domain knowledge-based security bug reports prediction \cite{PS14} \\

PS15 & Research on Feature Optimization Scheme Based on Data Feature Enhancement \cite{PS15} \\

PS16 & A process framework for the classification of security bug reports \cite{PS16} \\

PS17 & Automatically Identifying Bug Reports with Tactical Vulnerabilities by Deep Feature Learning \cite{PS17} \\

PS18 & Identifying security bug reports via text mining: An industrial case study \cite{PS18} \\

PS19 & Security bug reports classification using fasttext \cite{PS19} \\

PS20 & An improved text classification modelling approach to identify security messages in heterogeneous projects \cite{PS20} \\

PS21 & Leveraging Cross-Project Similarity for Data Augmentation and Security Bug Report Prediction \cite{PS21} \\

PS22 & Automatically identifying security bug reports via multitype features analysis \cite{PS22} \\

PS23 & Few-shot learning for security bug report identification \cite{PS23} \\

PS24 & SEDAC: A CVAE-Based Data Augmentation Method for Security Bug Report Identification \cite{PS24} \\

PS25 & Better security bug report classification via hyperparameter optimization \cite{PS25} \\

PS26 & Improving high-impact bug report prediction with combination of interactive machine learning and active learning \cite{PS26} \\

PS27 & Identifying Security Bug Reports Based Solely on Report Titles and Noisy Data \cite{PS27} \\

PS28 & Security Bug Report Prediction Within and Across Projects: A Comparative Study of BERT and Random Forest \cite{PS28} \\ \midrule

\multicolumn{2}{p{12.3cm}}{\textbf{Publications over the years:} 2010(1), 2014(1), 2017(1), 2018(2), 2019(3), 2020(2), 2021(5), 2022(4), 2023(2), 2024(3), 2025(4)} \\ \bottomrule

\end{tabular}
\end{table}

\section{Results and analysis} \label{sec:results}

This study provides a comprehensive map of various automatic techniques for identifying security bug reports. The complete map, which is too large to reproduce here, including metadata and other contextual information, can be found at \url{https://figshare.com/s/da2655372fe0494dfcef}.

The list of included primary studies is shown in Table \ref{tab:included-PSs}. We found primary studies from 2010 to 2025 on the topic. These studies are published across a diverse set of venues, including leading journals such as IEEE Transactions on Software Engineering, Information and Software Technology, and Software Quality Journal, as well as conferences like the Conference on Software Security and Reliability and the Conference on Mining Software Repositories.

In the following subsections, we answer the research questions listed in Table \ref{tab:research-questions}.

\subsection{RQ1: Which automatic techniques have been proposed?}

\begin{table}[!ht]
    \centering
    \scriptsize
    \caption{Automatic techniques proposed in the literature for identifying security bug reports}
    \label{tab:techniques}
    \begin{tabular}{p{1.9cm}p{3.5cm}p{4.7cm}p{0.9cm}}
    \toprule
    \textbf{Category} & \textbf{Technique} & \textbf{Relevant studies} & \textbf{Count}
    \\ \midrule

\multirow{8}{4cm}{Traditional \newline ML}
& Naive Bayes & PS1, PS2, PS4, PS6, PS7, PS9, PS10, PS11, PS12, PS15, PS16, PS17, PS20, PS26, PS27 & 15             \\ \cmidrule{2-4}

& Random Forest & PS1, PS3, PS5, PS6, PS7, PS9, PS11, PS12, PS13, PS17, PS20, PS23, PS25, PS28 & 14 
\\ \cmidrule{2-4}

& Logistic Regression & PS1, PS3, PS6, PS7, PS9, PS11, PS12, PS17, PS20, PS23, PS25, PS27 & 12
 \\ \cmidrule{2-4}

& Support Vector Machines (SVM) & PS3, PS5, PS7, PS8, PS9, PS11, PS20, PS22, PS23 & 9 
\\ \cmidrule{2-4}

& K-nearest Neighbors (KNN)  & PS6, PS9, PS11, PS12, PS17, PS20, PS25  & 7 
\\ \cmidrule{2-4}

& AdaBoost  & PS1, PS27 & 2
\\ \cmidrule{2-4}

& Decision Trees & PS1, PS12 & 2
\\ \cmidrule{2-4}

& Fasttext & PS19 & 1            
\\ \midrule

\multirow{5}{4cm}{Deep \newline learning}
& Multilayer Perceptron  & PS6, PS7, PS9, PS11, PS17, PS25 & 6
\\ \cmidrule{2-4}

& Convolutional Neural Networks & PS5, PS15, PS21 & 3 
\\ \cmidrule{2-4}

& Long Short-term Memory Networks  & PS11, PS17, PS21 & 3
\\ \cmidrule{2-4}

& Recurrent Neural Networks  & PS15 & 1
\\ \cmidrule{2-4}

& Gated Recurrent Units & PS25  & 1
\\ \midrule

BERT & BERT-based Model	& PS21, PS28 & 2
\\ \midrule

GPT & GPT4All-Falcon, Instruct, Open-Orca (Mistral), and Wizard	& PS3 & 1
\\ \midrule

Encoder -- decoder & CVAE-auto-encoder & PS24 & 1
\\ \midrule

Few shot learning & SetFit & PS23 & 1
\\ \midrule	

Active learning & --	& PS12, PS26 & 2
\\ \midrule

\multirow{2}{*}{Other} & Knowledge Graph with Cosine Similarity & PS14 & 1
\\ \cmidrule{2-4}

& SAS Text Miner & PS18 & 1
\\ \bottomrule

\end{tabular}
\end{table}

Table \ref{tab:techniques} presents an overview of the automatic techniques proposed in the literature to identify security bug reports. Traditional ML techniques remain dominant, with Naive Bayes emerging as the most frequently employed algorithm, appearing in fifteen studies. Random Forest is also widely used, with fourteen studies, followed by Logistic Regression with twelve studies. SVM and KNN are reported in nine and seven studies, respectively. Other traditional techniques, including AdaBoost and Decision Trees, show limited adoption, each appearing in two studies, while FastText is reported in a single study.
Deep learning techniques are comparatively less prevalent. Among them, the Multilayer Perceptron is the most commonly utilized, appearing in six studies. Convolutional Neural Networks and Long Short-term Memory Networks are each employed in three studies, whereas Recurrent Neural Networks and Gated Recurrent Units are each reported in one study.

Transformer-based approaches are less extensively explored in this context. BERT-based models are identified in two studies, while GPT-based models are reported in a single study. Similarly, encoder–decoder architectures, such as CVAE-based autoencoders, appear only once in the reviewed literature.

Emerging learning paradigms are also observed, albeit to a limited extent. A few-shot learning-based approach (prompt-free framework for few-shot fine-tuning of Sentence Transformers) was used in a single study (i.e., SetFit \cite{koch2015siamese}), while active learning was used in two studies. In addition, other approaches, including knowledge graph-based methods and tools such as SAS Text Miner, are reported in one study.

\subsection{RQ2: Which datasets have been used?}

We found that several datasets have been used in the literature to evaluate the proposed techniques for identifying security bug reports. Among them, Ambari\footnote{https://ambari.apache.org/} (20 times), Camel \footnote{https://camel.apache.org/} (20 times), Derby\footnote{https://db.apache.org/derby/} (19 times), Wicket\footnote{https://wicket.apache.org/} (19 times), and Chromium\footnote{https://www.chromium.org/chromium-projects/} (15 times) are the datasets most frequently utilized. Other datasets, such as those from Mozilla, RedHat, Ubuntu, and OpenStack, are used less frequently in the included primary studies. 

In general, the findings highlight the dominance of archived open-source datasets. In contrast, only three studies (PS18, PS20, and PS27) used closed-source data from proprietary companies, indicating a lack of evaluations of the proposed techniques in real industrial settings.

\subsection{RQ3: Which features have been used?}

Table \ref{tab:features-set} summarizes the features utilized in the literature to identify security bug reports. As shown in the table, the description of bug reports is the most frequently used feature, appearing in twenty-six studies. The summary field is the second most commonly used feature, reported in seventeen studies. In contrast, the title or heading of bug reports is used far less frequently, appearing in only two studies.
Other types of features are rarely considered. For example, meta features, such as reported time and priority, are identified in only one study. Overall, the findings indicate a strong reliance on textual features, particularly descriptions and summaries of bug reports.

\begin{table}[!ht]
    \centering
   \caption{Features used with techniques for identifying security bug reports}
     \label{tab:features-set}
    \begin{tabular}{p{4cm}p{6cm}p{1cm}}
    \toprule
    \textbf{Feature} & \textbf{Relevant studies} & \textbf{Count}
    \\ \midrule
Description & PS1, PS2, PS3, PS4, PS5, PS6, PS7, PS8, PS9, PS10, PS11, PS12, PS13, PS14, PS15, PS16, PS17, PS18, PS19, PS21, PS22, PS23, PS24, PS25, PS26, PS28 & 26 \\ \midrule
           
Summary & PS1, PS4, PS6, PS8, PS9, PS11, PS13, PS14, PS17, PS18, PS19, PS20, PS21, PS22, PS24, PS25, PS28 & 17 \\ \midrule

Title/Heading & PS23, PS27 & 2 \\ \midrule
        
Meta features, such as reported time, and priority & PS22 & 1 \\ \bottomrule

\end{tabular}
\end{table}

\subsection{RQ4: Which pre-processing techniques have been used?}
\begin{table}[!ht]
    \centering
    \caption{Pre-processing techniques (for textual features extraction) used in the literature for identifying security bug reports}
    \begin{tabular}{p{4.3cm}p{5.8cm}p{0.8cm}}
    \toprule
        \textbf{Pre-processing technique} & \textbf{Relevant studies} & \textbf{Count} \\ \midrule
        
        TF-IDF & PS1, PS2, PS3, PS4, PS6, PS7, PS8, PS9, PS10, PS11, PS13, PS16, PS20, PS23, PS25, PS26, PS27, PS28 & 18 \\ \midrule

        Word2Vec (Continuous bag of words) & PS9, PS11, PS15, PS22 & 4 \\ \midrule
        
        BERT-based tokenizer & PS21, PS24, PS28 & 3 \\ \midrule

        CountVectorizer & PS12, PS18, PS19 & 3 \\ \midrule

        Fasttext-based embedding & PS17 & 1 \\ \midrule

        MPNetTokenizer & PS23 & 1 \\ \midrule
        
        Custom Vectorization techniques & PS5 & 1 \\ \bottomrule
        
    \end{tabular}
    \label{tab:Pre-processing-techniques}
\end{table}

Table \ref{tab:Pre-processing-techniques} presents the pre-processing techniques (for textual features extraction) employed in the literature to identify security bug reports. As shown in the table, TF-IDF is the most widely used approach, appearing in eighteen studies, making it the predominant technique in this context.
Other vectorization and embedding methods are used less frequently. Word2Vec (Continuous Bag-of-Words) is reported in four studies, while both BERT-based tokenizers and CountVectorizer are used in three studies each.
The remaining techniques are rarely adopted. FastText-based embeddings, MPNetTokenizer, and custom vectorization approaches are each reported in a single study.

\subsection{RQ5: Which evaluation approaches have been used?}
Table \ref{tab:evaluation-approaches} presents the evaluation strategies used in the reviewed studies to identify security bug reports. The most commonly adopted approach is testing on a fixed set, which is applied in twenty studies, indicating a clear preference for this evaluation strategy. Cross-validation techniques are comparatively less prevalent. Specifically, 10-fold cross-validation is reported in three studies, while 5-fold cross-validation is used in two studies. Other cross-validation variants are rarely used; 3-fold cross-validation and repeated evaluation (20 repetitions) each appear in a single study.

\begin{table}[!ht]
    \centering
    \caption{Evaluation approaches used in the literature for identifying security bug reports}
    \begin{tabular}{p{4.8cm}p{6cm}p{0.9cm}}
    \toprule
        \textbf{Evaluation strategy} & \textbf{Relevant studies} & \textbf{Count} \\ \midrule
        Test on a fixed set & PS1, PS2, PS5, PS6, PS7, PS9, PS10, PS13, PS14, PS15, PS16, PS17, PS18, PS19, PS20, PS21, PS22, PS25, PS27, PS28 & 20 \\ \midrule
        10-fold cross-validation (CV) & PS4, PS12, PS26 & 3 \\ \midrule
        5-fold CV &  PS23, PS24 & 2 \\ \midrule
        20 repetitions & PS3 & 1 \\ \midrule
        3-fold CV  & PS11 & 1 \\ \bottomrule
    \end{tabular}
    \label{tab:evaluation-approaches}
\end{table}

\subsection{RQ6: Which evaluation metrics have been used?}

\begin{table}[!h]
    \centering
    \caption{Evaluation metrics used in the literature for identifying security bug reports}
    \begin{tabular}{p{2cm}p{8.8cm}p{0.8cm}}
    \toprule
        \textbf{Evaluation metrics} & \textbf{Relevant studies} & \textbf{Count} \\ \midrule
        Precision &  PS2, PS3, PS6, PS7, PS8, PS9, PS10, PS12, PS13, PS14, PS15, PS16, PS17, PS18, PS19, PS20, PS21, PS22, PS23, PS25, PS26, PS28 & 22 \\ \midrule
        
        F1-score & PS1, PS3, PS5, PS6, PS7, PS8, PS9, PS10, PS12, PS13, PS14, PS15, PS16, PS17, PS19, PS20, PS21, PS22, PS23, PS26, PS28 & 21 \\ \midrule
        
        Recall & PS3, PS7, PS8, PS9, PS10, PS12, PS13, PS14, PS15, PS16, PS17, PS18, PS19, PS20, PS21, PS22, PS23, PS25, PS26, PS28 & 20 \\ \midrule

        G-measure & PS1, PS6, PS9, PS11, PS14, PS15, PS19, PS20, PS24, PS25, PS28 & 11 \\ \midrule
        
        Accuracy & PS1, PS3, PS7, PS10, PS13, PS15, PS17, PS18, PS22, PS27 & 10 \\ \midrule
        
        AUC & PS1, PS4, PS7, PS12, PS15, PS20, PS21, PS23, PS26, PS27 & 9 \\ \midrule
        
        MCC & PS5, PS7, PS23 & 3 \\ \bottomrule
        
    \end{tabular}
    \label{tab:evaluation-metrics}
\end{table}
Table \ref{tab:evaluation-metrics} summarizes the evaluation metrics used in the literature to identify security bug reports. Precision is the most frequently used metric, appearing in twenty-two studies, followed closely by F1-score and Recall, which are reported in twenty-one and twenty studies, respectively.
The G-measure is also relatively common, appearing in eleven studies, while accuracy is used in ten studies. Area Under the Receiver Operating Characteristic Curve (AUC) is reported in nine studies, indicating moderate usage. In contrast, the Matthews Correlation Coefficient (MCC) is less frequently adopted, appearing in only three studies.

\section{Discussion}\label{sec:discussion}
\textbf{Prediction accuracy of the proposed automatic techniques for identifying security bug reports:} Our findings show that the literature on identifying security bug reports has evaluated a range of automatic techniques, including traditional ML and deep learning-based techniques, as well as more advanced LLMs. Traditional ML techniques, such as Logistic Regression and Random Forest, remain strong baselines [PS3, PS23, PS28, PS27]. By contrast, GPT- and BERT-based models have shown limited effectiveness in identifying security bug reports, as reported in PS3 and PS28. Additionally, PS23 found that SetFit, a few-shot learning approach, outperforms commonly used traditional ML techniques, including Logistic Regression, SVM, and Random Forest. However, the findings in these studies are difficult to compare directly because they rely on different experimental settings, datasets, and sets of techniques. For example, PS23 uses only traditional ML techniques and SetFit, whereas PS3 is limited to GPT-based models and traditional ML techniques. As a result, the literature does not yet provide a comprehensive comparative evaluation of GPT- and BERT-based models, traditional ML models, and SetFit in a consistent experimental setting.

\noindent\textbf{Data imbalance:} We found that the datasets used in the literature for evaluating the proposed techniques to identify security bug reports suffer from severe data imbalance. As shown in Table~\ref{tab:db-imbalance}, security bug reports constitute only a small fraction of the total reports in the most frequently used datasets. For example, Camel and Ambari each contain only 7.4\% and 5.6\% security bug reports, respectively, while Wicket contains just 4.7\%. The imbalance is even more pronounced in Chromium, where security bug reports account for only 1.92\% of the dataset, despite the dataset containing 41,940 bug reports in total. This indicates that the most widely used benchmarks are dominated by non-security bug reports, which can bias learning algorithms toward the majority class and make the accurate identification of security-related reports particularly challenging.

To address the issue of data imbalance in security bug reports datasets, several studies have applied data imbalance handling techniques, including Mahakil \cite{bennin2017mahakil} and the Synthetic Minority Over-sampling Technique (SMOTE) \cite{chawla2002smote}. Among the reported techniques, SMOTE is the most widely used, appearing in six studies [PS1, PS3, PS7, PS12, PS25, PS26]. PS1, PS12, and PS25 reported performance improvements when using SMOTE. In particular, PS25 emphasized that careful hyperparameter optimization of both learners and data pre-processors can improve the effectiveness of SMOTE for security bug report identification. PS7 further compared multiple imbalance-handling techniques, including SMOTE, Mahakil, and ROSE \cite{lunardon2014rose}, and reported ROSE as the most effective for addressing data imbalance in this context. 
Overall, the reviewed studies suggest that data imbalance is a central challenge in identifying security bug reports. Given that several widely used datasets contain fewer than 10\% security bug reports, future work should explicitly account for class imbalance during model training and evaluation. In particular, researchers should systematically compare imbalance-handling techniques, such as SMOTE, Mahakil, and ROSE, across consistent experimental settings to better understand their impact on prediction performance.

\begin{table}[ht]
    \centering
\caption{Data imbalance in the most widely used datasets in the literature on identifying security bug reports, Bug Reports (BRs)}
    \label{tab:db-imbalance}
    \begin{tabular}{ccccc}
    \toprule 
\textbf{Project} &  \textbf{Security BRs}  &  \textbf{Non-Security BRs} & \textbf{Total BRs} & \textbf{Count of usage}
      \\\midrule

Camel &  74 (7.4\%) & 926 & 1000 & 20 times \\

Ambari & 56 (5.6\%) & 944 & 1000 & 20 times  \\

Derby & 179 (17.9\%) & 821 & 1000 & 19 times  \\

Wicket & 47 (4.7\%) & 953 & 1000 & 19 times  \\

Chromium &  808 (1.92\%) & 41132 & 41940 & 15 times  \\
\bottomrule

\end{tabular}
\end{table}

\noindent\textbf{Lack of use of robust evaluation approaches and evaluation metrics:}
As shown in Table \ref{tab:db-imbalance}, the security bug reports datasets are often unbalanced. Therefore, it is important to use evaluation metrics that are robust to data imbalance, such as AUC and MCC, when assessing the proposed automatic techniques. However, we found that only 9 of 28 studies used AUC, and only 3 used MCC; see Table \ref{tab:evaluation-metrics}.

Similarly, we noticed that robust evaluation approaches, such as 10-fold CV, are lacking in the included primary studies. 20 out of 28 are based solely on a fixed test set to evaluate the performance of the proposed techniques. In contrast, only 7 studies employ cross-validation-based approaches, see Table \ref{tab:evaluation-approaches}.

\noindent\textbf{Lack of industrial validations and extensive use of archived open-source data:} Our findings show that the existing work on identifying security bug reports mainly uses archived open-source data to evaluate the proposed automatic techniques. Of 28 primary studies, only 3 used closed-source data from proprietary companies, suggesting a lack of evaluation of the proposed techniques in real industrial settings. Thus, future work should focus on evaluating these techniques in real industrial settings to assess their performance. In addition, we observed that the existing work is primary focused on prediction accuracy of the proposed automatic, other aspects such as the techniques generalizability, concept drift, and explainability have not been studied. Thus, future work should consider these aspects together with prediction accuracy.

\noindent\textbf{Cross-project prediction and its impact:} We observe that several studies [PS9, PS10, PS15, PS19, PS21, PS28] have investigated cross-project prediction for security bug report identification and have shown improvements in performance. However, the use of this approach remains relatively limited, with only 6 of the 28 reviewed studies using it. Given the scarcity of labeled security bug reports and the severe class imbalance in commonly used datasets, cross-project prediction presents a promising direction for future research. In particular, future studies should further examine whether knowledge learned from one or more source projects can improve prediction performance on target projects with limited labeled data.

\section{Conclusion}\label{sec:conclusion}
In this study, we presented a comprehensive systematic mapping of automated techniques for identifying security bug reports. The results show that the literature explores a wide range of techniques, including traditional ML and deep learning, as well as more advanced LLMs.

The findings further highlight key challenges faced in this area, particularly the limited availability of labeled data and the presence of class imbalance in datasets. In addition, most studies rely on archival data from open-source repositories, often including projects such as Wicket, Ambari, Camel, and Derby, while evaluations in real-world industrial settings remain scarce.

In general, the current literature on security bug report identification lacks comprehensive evaluations of the proposed techniques, making it difficult to draw reliable conclusions about their effectiveness. Future research should prioritize more rigorous and comparative evaluations, address data-related challenges, and investigate the applicability of proposed techniques in realistic industrial environments.

\bibliography{sn-bibliography}

\end{document}